\documentclass[%
 aip,
 jmp,%
 amsmath,amssymb,
 reprint,%
]{revtex4-1}

\usepackage{graphicx}
\usepackage{dcolumn}
\usepackage{bm}

\preprint{AIP/123-QED}

\setcitestyle{numbers,square}

\usepackage{braket}
\usepackage{graphicx}
\usepackage{dcolumn}
\usepackage{bm}
\usepackage[normalem]{ulem} 
\usepackage{xcolor}

\usepackage{hyperref}
\hypersetup{
    colorlinks,%
    citecolor=blue,%
    linkcolor=blue,%
    urlcolor=blue
}

\newcommand{\orcid}[1]{\href{https://orcid.org/#1}{\includegraphics[width=8pt]{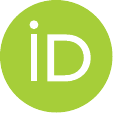}}}

\begin{document}

\title{Resonant helical multi-edge transport in Sierpi\'nski carpets}

\author{M. A. Toloza Sandoval\orcid{0000-0002-1071-6665}}
\email{marcelo.sandoval@lnnano.cnpem.br}
\affiliation{Ilum School of Science, Brazilian Center for Research in Energy and Materials (CNPEM), Campinas, SP, Brazil.}

\author{A. L. Araújo\orcid{0000-0002-6835-6113}}
\email{augusto.araujo@ilum.cnpem.br}
\affiliation{Ilum School of Science, Brazilian Center for Research in Energy and Materials (CNPEM), Campinas, SP, Brazil.}

\author{F. Crasto de Lima\orcid{0000-0002-2937-2620}} 
\email{felipe.lima@ilum.cnpem.br}
\affiliation{Ilum School of Science, Brazilian Center for Research in Energy and Materials (CNPEM), Campinas, SP, Brazil.}

\author{A. Fazzio\orcid{0000-0001-5384-7676}}
\email{adalberto.fazzio@ilum.cnpem.br}
\affiliation{Ilum School of Science, Brazilian Center for Research in Energy and Materials (CNPEM), Campinas, SP, Brazil.}

\date{\today}

\begin{abstract}

{In recent years, synthesis and experimental research of fractalized materials has evolved in a paradigmatic crossover with topological phases of matter. We present here a theoretical investigation of the helical edge transport in Sierpinski carpets (SCs), combining the Bernevig-Hughes-Zhang (BHZ) model and the Landauer approach. Starting from a pristine two-dimensional topological insulator (2DTI), according to the BHZ model, our results reveal resonant transport modes when the SC fractal generation reaches the same scale as the space discretization; these modes are analyzed within a contour plot mapping of the local spin-polarized currents, shown spanned and assisted by inner-edge channels. From such a deeply fractalized SC building block, we introduce a rich tapestry formed by superior SC hierarchies, enlightening intricate patterns and unique fingerprints that offer valuable insights into how helical edge transport occurs in these fractal dimensions.
}


\end{abstract} 

\maketitle

Fractals are intriguing mathematical objects that present complex patterns characterized by self-similarity; they exhibit a similar structure at different scales, found in abundant examples, from turbulent flows to crystal formation, encoding nature's hidden order \cite{Mandel, Vic, Bun}. In condensed matter physics and transport phenomena, the role of fractality can be pivotal and has garnered much attention in recent years \cite{Man23, Jun, Man22, Bie, Xu, Ili, Fre, Pai, Kem, Vee17, Sti, Vee}. Groundbreaking advances include the synthesis and experimental research of fractalized materials \cite{Bie, Xu, Kem}, which bring a demand to explore the crossroad between fractal branches and topological phases of matter \cite{Man23, Jun, Man22, Ili, Fre, Pai, Vee17}, within a paradigmatic research roadmap with a large venue for fundamental and applied research. 
        
One of the fundamental properties of two-dimensional (2D) topological insulators (TIs) is the robust transport signature due to the time-reversal symmetry-protected helical edge channels \cite{Has, Qi}. However, different mechanisms have been shown able to blur such a signature, producing localization and coherent backscattering.
Vacancies and other defects can mediate the inter-edge hybridization \cite{Tiw,Pez}, reshape the local currents \cite{Dan}, or lead to quantum percolation \cite{Chu}. In fractalized TIs, a central mechanism is enabled when the fractal generation produces interacting inner edges and hybridization as a consequence \cite{Son}; when the fractal generation reaches the level of individual sample sites, that is, the same scale as the space discretization, the transport properties will dramatically change. This exotic template then generates a rich and unexplored tapestry formed by higher fractal generations that start from the (afore-described) deeply fractalized 2D TI. 

In this paper, our systematic investigation reveals the emergence of transport modes with unique fingerprints, providing a transparent view of the topological matter's behavior when submitted to a fractal generation process that introduces interacting inner edges. Such an intricate question is theoretically addressed by looking for the helical edge transport in Sierpinski carpets (SCs) \footnote{Plane fractal with Hausdorff dimension $d_f=ln(8)/ln(3)$.}, using the Bernevig-Hughes-Zhang (BHZ) model \cite{Ber} together with the Landauer approach \cite{Lan57,Lan70} within the Kwant package \cite{Gro}. 

\begin{figure}[h]
\includegraphics[width=\columnwidth]{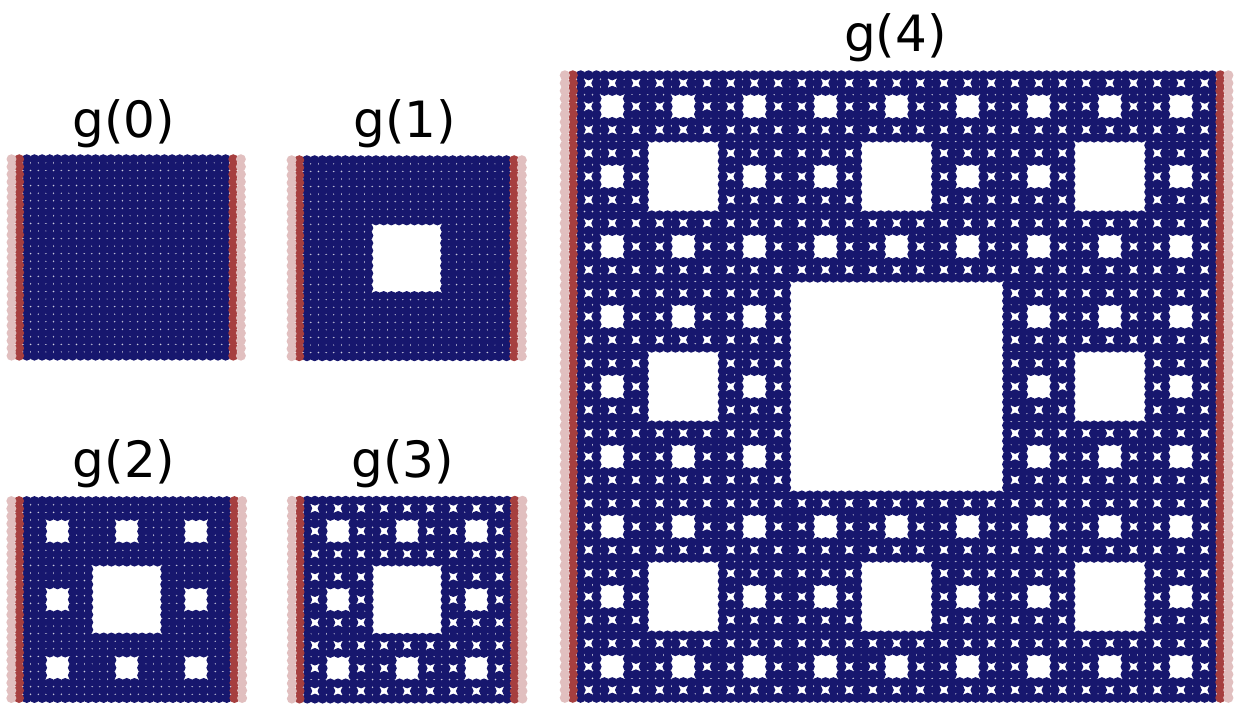}
    \caption{\label{fig:fig1} Transport geometries. Red sites indicate the left and right leads, while blue sites indicate the scattering region fractalized to form different generations of SCs (g(0), g(1), g(2), g(3), and g(4)). The complete information is provided in the text.}
\end{figure}

\begin{figure*}[t]
\includegraphics[width=18.0cm]{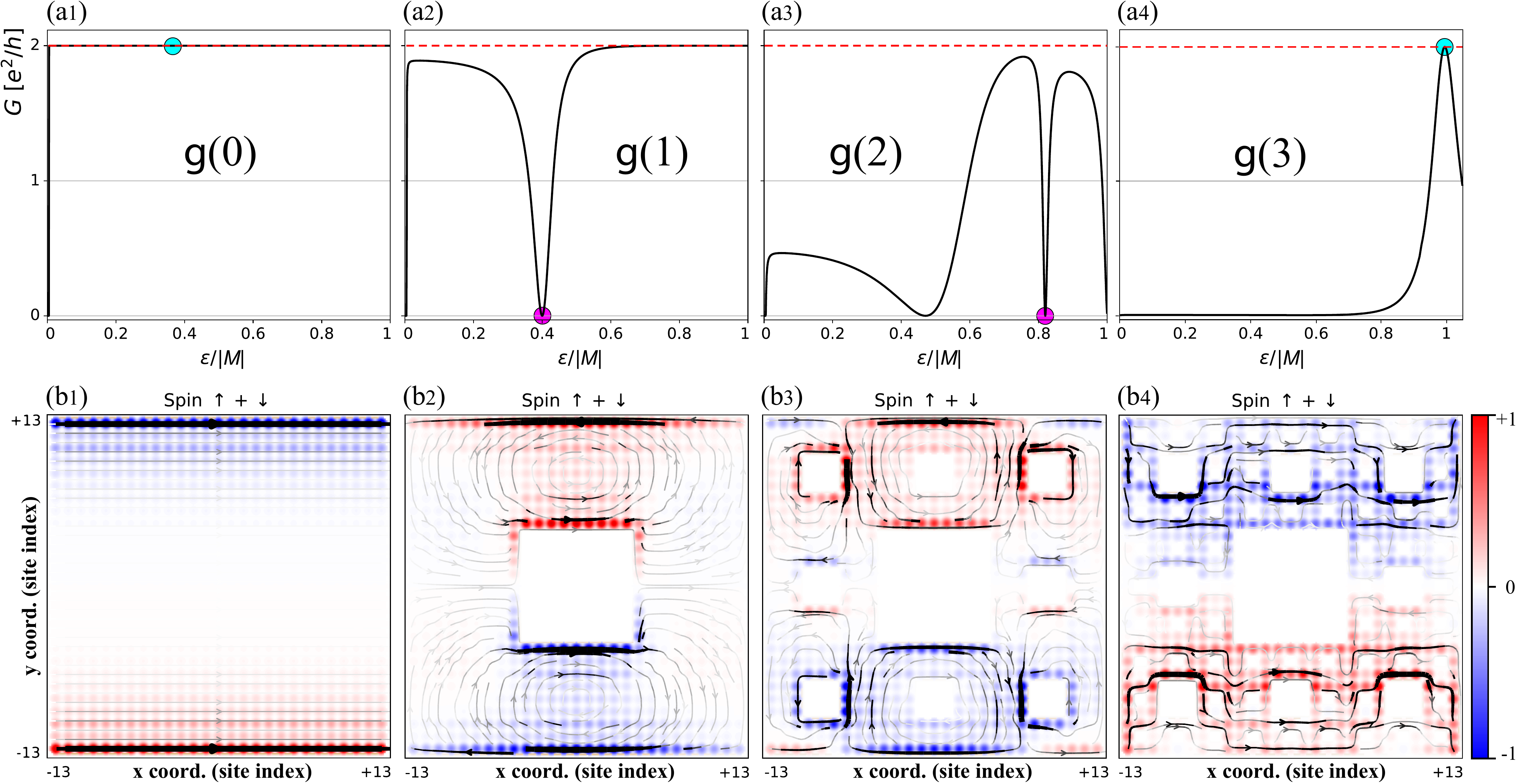}
\caption{\label{fig:fig2} Panels from left to right correspond to increasing fractal generations as labeled. The graphics (a1)-(a4) show the energy dependence of the in-gap conductance, with energy ranging in units of $|M|$, setting the  bulk bandgap from -1 to 1, noting that $G(-E)=G(E)$, i.e. particle-hole symmetry. The diagrams (b1)-(b4) present the local current density represented by black lines and arrows (with magnitude, direction, and sense) and the $z$-component of the spin polarization given according to the color barcode; for each generation, the color dot at the superior panel indicates the projected state at the lower panel.
} 
\end{figure*}

\begin{figure*}[t]
\includegraphics[width=18.0cm]{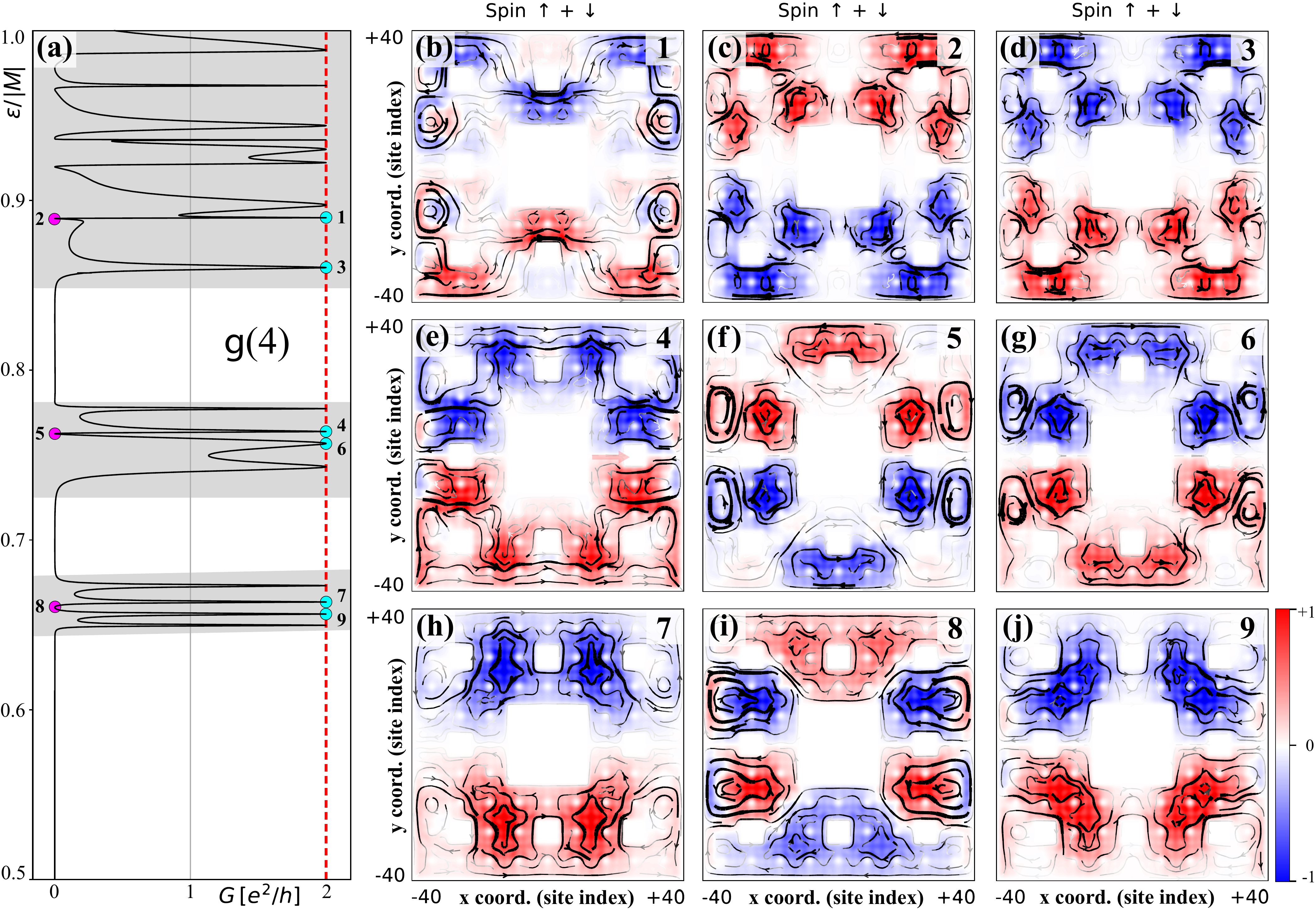}
\caption{\label{fig:fig3} The transport setup presents a 4th-generation g(4) SC as a scattering center comporting 4096 sites, with homogeneously attached semi-infinite leads formed by ideal 2DTIs (as shown in Fig.~\ref{fig:fig1}). (a) Energy-dependent in-gap conductance displayed in a miniband structure of resonant states and well-defined sub-gaps (white regions). (b)-(j) local current densities for the numbered energies in (a); local current densities are given by black lines and arrows and $z$-component of the spin polarization according to the color barcode.}
\end{figure*}


To study the helical edge transport characteristic of a 2D TI, we use the $4\times 4$ BHZ Hamiltonian written in a convenient dimensionless form \cite{Mac, Suk}, 
 \begin{eqnarray}
\mathcal{H}_{eff}(\mathbf{k}_{\parallel}) =
\left[\begin{array}[c]{c c}
  H(\mathbf{k}_{\parallel}) & 0  \\\\
  0 & H^*(-\mathbf{k}_{\parallel})
\end{array}\right]
\end{eqnarray}
with
\begin{equation}
H(\mathbf{k}_{\parallel})=\mathbf{k}_{\parallel}\cdot {\boldsymbol{\tau}} + \left(\tilde{M}+\mathcal{W} k_{\parallel}\right)\tau_z
\end{equation}
where $\mathbf{k}_{\parallel}=(k_x,k_y)$, $\boldsymbol{\tau}=(\tau_x,\tau_y,\tau_z)$ is the Pauli vector that operates on the orbital indices, $\tilde{M}=M/|M|$, $\mathcal{W}=|M|\,B/A^2$ is the Wilson mass \cite{Mes,Ara}, $A$, $B$ and $M$ are the BZH-model parameters \cite{Ber, Qi} in the particle-hole symmetric configuration.  

We use this model to describe a transport setup comporting leads and the scattering center formed by 2D TIs based on HgTe quantum wells ($A=3.65$\,eV{\AA}, $B=-68.6$\,eV{\AA}$^2$, and $M=-10$\,meV), which is numerically regularized in a square lattice to calculate the wave functions, conductance, local currents and spin polarization by performing the Kwant toolkit \cite{Gro}.          

Matching wave function across the transport setup, such a toolkit allows computing the scattering matrix $S_{ij}$ corresponding to an incoming propagating mode $i$ (into the scattering center) and an outgoing $j$. In a two-terminal transport setup with left $L$ and right $R$ leads, the conductance $G(E)$ is then obtained by employing the Landauer formula,
\begin{equation}
G(E) = G_0\hspace{-.1in}\sum_{i\in R, \, j\in L}\hspace{-.1in}|S_{ij}(E)|,
\end{equation}
where $G_0=2\,e^2/h$ is the quantized (in-gap) conductance characteristic of 2DTIs, i.e., $G(E) = G_0$ defines the pristine quantum spin Hall conductance \cite{Pez, Focassio_2021}.

We recall central features of the helical edge states in a usual 2D TI: they are extended along the edge (or longitudinal direction), whereas the penetration depth determines the decaying of the wavefunction along the transverse direction. Within the fractal generation procedure, the zero-order iteration (i.e., g(0) in Fig.~\ref{fig:fig1}) corresponds to a pristine 2D TI, which forms both leads and the scattering center used in the transport calculation setup. Considering a scattering region as a square sample comporting 729 sites (i.e., 27$\times$27) with attached semi-infinite electrodes that present the same width as the scattering center, we show in Fig.~\ref{fig:fig2}(a1) the obtained conductance between left and right leads, ranging the energy within the bulk gap; the lower panel [Fig.~\ref{fig:fig2}(b1)] shows the corresponding helical edge modes, with the current density represented by black arrows (and lines), and the $z$-component of the spin polarization given according to the color barcode.

The first step to form a SC starting from the pristine 2D TI is to remove a square with 81 sites at the system's center, resulting in a first-generation SC comporting 648 sites, labeled by g(1) in Fig.~\ref{fig:fig1}. Here, the overlap between states at opposite edges, with the same spin component and opposite velocities, is enabled by the internal edge channels formed between the 2D bulk and the hollows generated by the fractal procedure. When the distance between opposite edges (including internal and external edges) is in the same order as the penetration depth, such an overlap leads to the appearance of an antiresonance, i.e., a specific energy for which the conductance vanishes. It is shown in Fig.~\ref{fig:fig2}(a2), with circulating local currents displayed in detached vortices, corresponding to the strongly localized state projected in Fig.~\ref{fig:fig2}(b2). Understanding how this initial step affects the transport properties is fundamental to analyzing the next steps of the fractalization process. Within the 2nd-generation SC, the increasing number of the interacting edge channels enhances the hybridization, starting the formation of a well-defined sub-gap around $\varepsilon /|M|\approx 0.5$, shown in Fig.~\ref{fig:fig2}(a3), accompanied by the strongly localized anti-resonant state mapped in Fig.~\ref{fig:fig2}(b3). Furthermore, note that this system does not support states that preserve the pristine conductance $G_0$.

In contrast with the earlier SC generation, the setup comprising the 3rd generation SC (i.e., g(3) in Fig.~\ref{fig:fig1}) exhibits a resonant conductance, shown in Fig.~\ref{fig:fig2}(a4), due to constructive interference between multiple conducting edge channels, giving rise to the extended state projected in Fig.~\ref{fig:fig2}(b4). It can be seen within a quantum percolation picture by considering the spatial arrangement between edges and sub-edges. Although the penetration depth is enough to induce transverse percolation channels (allowing backscattering), the resonant state persists. Here, a current channel between source and drain is ruled by a peculiar alignment between longitudinal percolation channels spanned along multiple edges.
            
We use the self-similarity property to construct the 4th generation SC, i.e., g(4), starting from g(3) as a building block or unity, as shown in Fig.~\ref{fig:fig1}. This complex structure inherits the g(3) fingerprints within a hierarchical fractal ordering; indeed, the obtained in-gap conductance shows resonant modes displayed in a miniband structure, with forbidden sub-gaps and sub-bands (grey regions) with accessible states. The numbered resonant and anti-resonant states are projected: for the resonant point with $G(E)=2$, the pristine transport signature, that is, bipartition of the spin-current with upper edge being spin down ($S_z = -1$) and botton edge spin up ($S_z = 1$). For the antiresonant point the bipartition present the different spin current, for instance see Fig.~\ref{fig:fig3}(i). Here given the spin-momentum locking in TI \cite{PhysRevLett.95.226801}, such inverted spin polarization characterize contra propagation of the electrons.

In conclusion, our systematic theoretical investigation of the helical edge transport phenomena in SCs revealed the emergence of transport patterns with exclusive signatures, providing a deeper insightful view of a TI submitted to the fractalization process that produces interacting inner edges. Undergoing successive generations, starting from the deeply fractalized building block g(3), that comports an unusual resonant state, a tapestry comprising superior SC hierarchies is introduced and meticulously examined; for g(4), the calculated in-gap conductance shows resonant modes displayed in a miniband structure, as a fingerprint underlying the fractal ordering and self-similarity. Our findings expand our understanding of the fundamental principles within this complex intersection between topological phases of matter and fractality, as well as offer a fertile ground for tailoring the transport properties of TIs through controlled fractalization within the profuse and nuanced behaviors exhibited by fractalized materials. 

\section*{acknowledgments}
The authors acknowledge financial support from
the Brazilian agencies CNPq (INCT - Materials
Informatics, INCT - Nanocarbono, and SisNANO - project 442493/2019-3), FAPESP (grants 2017/02317-2, 22/08478-6, and 2023/12336-5, FINEP (MQSM).

\section*{Data availability}
The data supporting the findings of this study are available within the article; additional data are available from the authors upon reasonable request.

\bibliography{refs-aux}
\end{document}